%% file: vlasov_renorm_transfer_newfigs.tex
\documentclass[aps,pre,showpacs,showkeys,amsmath,amsfonts,amssymb,notitlepage,superscriptaddress,nofootinbib]{revtex4-1} 
\usepackage[dvips]{graphicx}
\usepackage[applemac]{inputenc}
\usepackage{bbm}
\usepackage{xcolor}
\usepackage[colorlinks,linkcolor=blue,anchorcolor=blue,citecolor=blue,filecolor=blue,menucolor=blue,urlcolor=blue]{hyperref}
\usepackage{breakurl}

\newcommand{\beq}{\begin{equation}} 
\newcommand{\eeq}{\end{equation}}
\newcommand{\bea}{\begin{eqnarray}} 
\newcommand{\eea}{\end{eqnarray}}

\newcommand{\mpd}[3]{\frac{\partial^{#3} #1}{\partial {#2}^{#3}}}
\newcommand{\pd}[2]{\frac{\partial #1}{\partial #2}}
\newcommand{\td}[2]{\frac{d #1}{d #2}}

\newcommand{\me}[1]{\left\langle #1 \right\rangle }
\newcommand{\pb}[1]{\left\lbrace #1 \right\rbrace}

\input epsf 
 
\begin{document} 
 
\title{Coarse-grained collisionless dynamics with long-range interactions} 

\author{Guido Giachetti} 
\email{ggiachet@sissa.it} 
\affiliation{SISSA, via Bonomea 265, I-34136 Trieste, Italy}
\author{Alessandro Santini}
\email{alessandro.santini2@stud.unifi.it} 
\affiliation{Dipartimento di Fisica e Astronomia, Università di Firenze, via G.\ Sansone 1, I-50019 Sesto Fiorentino, Italy}  
\author{Lapo Casetti}
\email{lapo.casetti@unifi.it} 
\affiliation{Dipartimento di Fisica e Astronomia, Università di Firenze, via G.\ Sansone 1, I-50019 Sesto Fiorentino, Italy}  
\affiliation{INFN, Sezione di Firenze, and INAF-Osservatorio Astrofisico di Arcetri, Firenze, Italy}  

\date{\today} 
 
\begin{abstract} 
We present an effective evolution equation for a coarse-grained distribution function of a long-range-interacting system preserving the symplectic structure of the non-collisional Boltzmann, or Vlasov, equation. We first derive a general form of such an equation based on symmetry considerations only. Then, we explicitly derive the equation for one-dimensional systems, finding that it has the form predicted on general grounds. Finally, we use such an equation to predict the dependence of the damping times on the coarse-graining scale and numerically check it for some one-dimensional models, including the Hamiltonian Mean Field (HMF) model, a scalar field with quartic interaction, a 1-$d$ self-gravitating system, and the Self-Gravitating Ring (SGR). 
\end{abstract} 
 
 
 
\maketitle 

\section{Introduction} 
\label{sec_intro}
Long-range interactions, whose potential energy decays with the distance $r$ between the interacting bodies slower than $r^{-d}$, where $d$ is the dimension of space  \cite{CampaEtAl:book,CampaEtAl:physrep}, are relevant to astrophysics and plasma physics \cite{BinneyTremaine:book,Nicholson:book}, since gravitational and unscreened Coulomb forces are long-ranged, as well as to condensed matter, given that dipolar interactions in $d=3$ or effective interactions between cold atoms in an electromagnetic cavity \cite{SchutzMorigi:prl2014,njp2016} are long-ranged, and occur also in two-dimensional fluids \cite{BouchetVenaille:physrep2012}. Systems with long-range interactions exhibit peculiar features both at equilibrium and out of equilibrium \cite{CampaEtAl:book,CampaEtAl:physrep,LevinEtAlphysrep:2014,prl2015}. In systems with long-range interactions, the dynamics is dominated by collective effects, rather than by binary collisions; as a consequence, the relaxation time towards equilibrium $\tau_{\text{R}}$ diverges with the number of particles $N$ \cite{Chandrasekhar:apj1941,BinneyTremaine:book}. In the $N \to \infty$ limit, or for times $t < \tau_{\text{R}}$ for finite $N$, the dynamics obeys the non-collisional Boltzmann, or Vlasov, equation \cite{BinneyTremaine:book,CampaEtAl:book,CampaEtAl:physrep}. Introducing the single-particle Hamiltonian for $N$ particles (for simplicity, we assume identical particles with unit masses),
\beq
H = \frac{p^2}{2} + U(q,t)\,, 
\label{H:single}
\eeq
where $q = (q_1,\ldots,q_d)$, $p = (p_1,\ldots,p_d)$, $U$ is the self-consistent potential
\beq
U(q,t) = \int dp'\, dq'\, f(q',p',t) V(|q - q'|) \,,
\eeq
$f(q,p,t)$ is the single-particle distribution function, $V(r)$ is the potential energy between two particles at distance $r$, we can write the Vlasov equation as 
\beq
\pd{f}{t} = \pb{H,f} \,,
\label{eq:vlasov}
\eeq
where $\pb{\cdot,\cdot}$ is the Poisson bracket, thus making explicit its symplectic structure.  This has important consequences \cite{Morrison:pla1980,PerezAly:mnras1996,Kandrup:apj1998}, e.g., the Vlasov equation is time-reversal-invariant and its dynamics is constrained by an infinite number of conservation laws: the Casimirs
\beq
\mathcal{C}[f] = \int dq\, dp\, C(f)
\eeq
are conserved for any choice of $C(f)$. Remarkably, also the Boltzmann entropy is a Casimir, corresponding to $C(f) = - f \ln f$, so that it is a constant of motion and no $H$ theorem holds. All these properties seem to suggest that no relaxational dynamics is possibile: any time dependence of $f$ should survive forever in the $N\to\infty$ limit, and at least up to $t \approx \tau_{\text{R}}$ when collisional effects set in for a large but finite system. Numerical results depict a totally different scenario: starting from a generic initial condition, a given observable exhibits oscillations that damp out on a rather fast time scale not depending on $N$ (at variance with $\tau_{\text{R}}$) until it attains a nearly constant value. The paradigmatic example is gravitational collapse \cite{Henon:1964AnnAp1964,vanAlbada:mnras1982,Sylos:mnras2012}, where the relevant observable is either the gravitational radius or the virial ratio, so that the damped oscillations are termed ``virial oscillations'', and this noncollisional relaxation is referred to as ``violent relaxation'' \cite{Lynden-Bell:mnras1967}. Violent relaxation  is a universal phenomenon, occurring in any long-range-interacting system; the state reached after violent relaxation  is referred to as a quasi-stationary state (QSS), may be very far from thermal equilibrium \cite{CampaEtAl:book,epjb2014,prerap2015,mnras2018,jstat2019}, and in a finite system will eventually relax to equilibrium for $t > \tau_{\text{R}}$. Despite many advances \cite{CampaEtAl:book,CampaEtAl:physrep,LevinEtAlphysrep:2014} a theory able to predict these states given a generic initial condition is still missing. It is widely believed that the mechanism of violent relaxation  is similar to Landau damping \cite{Kandrup:apj1998,BarreOlivettiYamaguchi:jstat2010,BarreOlivettiYamaguchi:jphysa2011}. Basically this means that the Vlasov dynamics does never actually stop: rather it trickles down towards smaller and smaller scales until it no longer affects the behavior of any coarse-grained observable. Indeed, given a coarse-grained distribution function $\tilde{f}$, obtained by averaging $f$ over some finite volume $\Delta\Gamma$ in phase space, and any convex function $C(x)$, the corresponding Casimir $\mathcal{C}[\tilde{f}]$ decreases in time \cite{TremaineHenonLynden-Bell:mnras1986}. Despite this, a convincing quantitative picture of this process is still missing: our aim is then to contribute to filling this gap by providing an effective evolution equation for $\tilde{f}$. 

The paper is organized as follows. In Sec.\ \ref{sec_sympl_CG} we propose a general form of the effective evolution equation, up to coefficients, based on symmetry considerations only. Then, in Sec.\ \ref{sec_onedim} we explicitly perform the coarse graining and derive the complete equation in the one-dimensional case. Sec.\ \ref{sec_numerics} is devoted to predicting the dependence of damping times on the coarse-graining scale and checking the results against numerical simulations of some one-dimensional models: HMF model, a scalar field with quartic interaction, a 1-$d$ self-gravitating system, and the Self-Gravitating Ring (SGR). Finally, in Sec.\ \ref{sec_conclusions} we comment on the results we have obtained, discuss their relation to other approaches, open problems and future developments. To ease the reading, some proofs and some further details on the numerics are reported in appendices.
 
\section{Symplectic coarse graining} 
\label{sec_sympl_CG}
Many properties of an effective evolution equation for $\tilde{f}$ can be derived from symmetry considerations and very general assumptions, that define what we will refer to as \textit{symplectic coarse graining}. First of all,  if $\tilde{f}$ is normalized to unity then a coarse-grained single-particle Hamiltonian $H[\tilde{f}]$ is defined as in Eq.\ \eqref{H:single}, with $\tilde{f}$ in place of $f$; to ease the notation, we simply write $H$ in place of $H[\tilde{f}]$. We then assume that the coarse graining procedure does not depend on the choice of the canonical coordinates, preserving the symplectic structure; therefore, the dynamical evolution of $\tilde{f}$ can be expressed in terms of Poisson brackets. Moreover, we assume that Poisson brackets contain functions of $H$ and $\tilde{f}$ alone and are linear in $\tilde{f}$; physically, this means that particles interact only via $H$, as in the Vlasov equation \eqref{eq:vlasov}. These assumptions imply that
\beq
\pd{\tilde{f}}{t} = \mathcal{L}_{H}(\tilde{f}) \,,
\eeq 
where $\mathcal{L}_{H}(\tilde{f})$ depends on $H$, acts linearly on $\tilde{f}$, and its most general form is a linear combination of nested Poisson brackets where $\tilde{f}$ appears only once, that is, 
of terms of the form $\left\{\lambda_1(H),\left\{\lambda_2(H),\left\{\cdots \left\{\lambda_k(H),\tilde{f} \right\}\cdots \right\} \right\}\right\}$,
where the $\lambda_k$'s are generic functions of $H$.
By repeatedly using the identity $\pb{\lambda_k(H), \cdot } = \lambda_k^{\prime}(H) \pb{H, \cdot}$ and denoting with $\pb{H, \cdot}^{n} \tilde{f}$ the $n$ nested Poisson brackets, i.e.,
\beq
\pb{H, \cdot}^{n} \tilde{f} = \underbrace{\left\{H,\left\{H,\left\{\cdots \left\{H,\tilde{f} \right\}\cdots \right\} \right\}\right\}}_{n~\text{times}}\,,
 \eeq 
we can thus write
\begin{equation} 
\label{Mostgeneral2}
\pd{\tilde{f}}{t} =  \pb{H, \tilde{f}} + \sum^{\infty}_{n=2} \mu_n (H) 
\pb{H, \cdot}^{n} \tilde{f}\,,
\end{equation}
where the $\mu_n$'s are generic functions that absorb the coefficients of the linear combination. In Eq.\ \eqref{Mostgeneral2} we have highlighted the first term of the sum, assuming $\mu_1(H) \equiv 1$, as is reasonable since $\tilde{f} \to f$ and Eq.\ \eqref{Mostgeneral2} must reduce to Eq.\ \eqref{eq:vlasov} when\footnote{We are implicitly assuming that the sum of the contributions to Eq.\ \eqref{Mostgeneral2} with $n > 2$ vanishes when $\Delta\Gamma \to 0$.} 
$\Delta\Gamma \to 0$.
We note that both the normalization of $\tilde{f}$ and the total energy $E[\tilde{f}]$ are conserved by Eq.\ \eqref{Mostgeneral2}, as required by a physically sound evolution (see Appendix \ref{app_conservationlaws}). 

Equation \eqref{Mostgeneral2} is the most general outcome of symplectic coarse graining. The terms of the sum on the r.h.s.\ of Eq.\ \eqref{Mostgeneral2} containing an odd number of Poisson brackets do not break time-reversal invariance, so that they renormalize the time-reversible Vlasov evolution, while those containing an even number of brackets break the time-reversal invariance and may account for dissipation.  However, we expect that not all the  possible $\mu_n$'s are physically admissible. For instance, as already mentioned, all the convex Casimirs defined by the coarse-grained distribution function $\tilde{f}$ must decrease with time. It is not easy to impose such a constraint on Eq.\ \eqref{Mostgeneral2}, but the lowest-order truncation of the latter equation, obtained by setting $\mu_n = 0$ $\forall\, n > 2$,
\beq
\pd{\tilde{f}}{t} = \pb{H,\tilde{f}} + \mu_2(H) \pb{H,\pb{H,\tilde{f}}}\,,
\label{eq:general_lo}
\eeq
with the additional constraint $\mu_2(x) > 0$ $\forall\, x$, does satisfy this  constraint (see Appendix \ref{app_convex_general}), and actually describes a Vlasov-like evolution with added diffusive effects, hence admitting a relaxational behavior. Indeed, $\pb{H,\cdot}$ is proportional to the directional derivative along the Hamiltonian flow generated by $H$, so that $\pb{H,\pb{H,\cdot}}$ is a sort of anisotropic Laplacian and the second term on the r.h.s.\ of Eq.\ \eqref{eq:general_lo} describes a diffusion taking place along the Hamiltonian flow, whose strength depends on $\mu_2$, that in turn will depend on the coarse-graining scale $\Delta\Gamma$; this will become apparent in the one-dimensional case that we are going to tackle in the following. Once $\Delta\Gamma$ is fixed, Eqs.\ \eqref{Mostgeneral2} and \eqref{eq:general_lo} are expected to be appropriate to describe the evolution of an observable which is not sensitive to the structure of $f$ on scales smaller than $\Delta\Gamma$ itself. Choosing as $\Delta\Gamma$ the smallest scale the observable of interest is sensitive to, the odd (conservative) terms in Eqs.\ \eqref{Mostgeneral2} and \eqref{eq:general_lo} will eventually relocate the dynamics on scales smaller than $\Delta\Gamma$, while the even (dissipative) terms will  erase such information in $\tilde{f}$, thus effectively describing the dynamics of the chosen observable. 

\section{Effective equation for one-dimensional systems} 
\label{sec_onedim}
Let us perform a symplectic coarse graining  and obtain an explicit evolution equation for the coarse-grained $\tilde{f}$ in the case of 1-$d$ systems, bounded in space or with periodic boundary conditions. In this case $H$ has one degree of freedom, so that, at a given time, it is integrable and a canonical transformation $(p,q) \mapsto (J,\vartheta)$ exists, where $(J,\vartheta)$ are action-angle variables. Being $H$ time-dependent in general, such a transformation leads to a Hamiltonian independent of the angle $\vartheta$ only at a given time $t$. The instantaneous flow will be such as to keep $J$ constant, and the angle will linearly evolve in time, 
\beq
\vartheta(t+\Delta t) = \vartheta(t) + \omega(J) \Delta t\,, 
\eeq
where $\omega(J) = dH/dJ$. Let us consider a (small) interval of actions $\Delta J = J_2 - J_1$, define $\overline{\omega}$ as the frequency $\omega$ averaged over $\Delta J$,
\beq
\overline{\omega} = \frac{1}{\Delta J} \int_{J_1}^{J_2} \omega(J')\, dJ' 
\eeq  
and consequently $\delta\omega = \omega - \overline\omega$, and a distribution function coarse-grained along $J$ as
\beq
\bar{f}(J,\vartheta) = \frac{1}{\Delta J} \int_{J_1}^{J_2} f(J',\vartheta)\, dJ'\,, 
\label{f_avJ}
\eeq
where $J$ is such that $\overline{\omega} = \omega(J)$. To get a truly coarse-grained distribution function one should average also over an interval of angles $\Delta\vartheta$, but it is more convenient to consider such an average as carried over a time interval $\Delta t$, that is, to assume we are blind to changes of the coordinates of the particles occurring on time scales smaller than $\Delta t$. This means that we neglect the time dependence of $H$ on a time scale  $\Delta t$, and we can use action-angle coordinates for times between $t$ and $t + \Delta t$, defining a non-constant coarse-graining scale on $\vartheta$, namely $\Delta \vartheta = \overline{\omega}\Delta t$. Our coarse-grained distribution function $\tilde{f}$ is then the function $\bar{f}$ given by Eq.\ \eqref{f_avJ}, further averaged over an interval of angles of width $\Delta\vartheta$ centered in $\vartheta$. As a consequence, we are not able to distinguish any change of $\tilde{f}$ on scales smaller than  $\Delta\Gamma = \Delta\vartheta\Delta J$. For times between $t$ and $t + \Delta t$ we have approximated the flow in phase space with a stationary one, so that its evolution operator should be written as 
\beq
U_{\Delta t} = e^{\Delta t \{H,\cdot \}}\,.
\eeq 
The latter is not constant over $\Delta\Gamma$, but within such volume we can consider $J$ and $\vartheta$ as uniformly distributed random variables, so that (up to very unlikely initial conditions) we can write
\beq
\tilde{f}_{t+\Delta t} = \tilde{U}_{\Delta t} \tilde{f}_{t}\,,
\label{eq:evolution1d}
\eeq
where we have replaced the evolution operator $U_{\Delta t}$ with the coarse-grained one 
\beq
\tilde{U}_{\Delta t} = \left\langle e^{\Delta t \{H,\cdot \} }\right\rangle_{\Delta\Gamma}~.
\label{eq:evol_operator}
\eeq 
The evolution dictated by Eq.\ \eqref{eq:evol_operator} satisfies the constraint on the evolution of convex Casimirs (see Appendix \ref{sec:convex1d}) and can be translated into a differential equation for the coarse-grained distribution function $\tilde{f}(t)$. 
To derive such an equation, we start by writing the evolution operator in action-angle variables,
\beq
\tilde{U}_{\Delta t} = \me{e^{ \Delta t \pb{H,\cdot}}}_{\Delta \Gamma} =  \me{e^{ - \omega(J^{\prime}) \Delta t \partial_{\vartheta}}}_{\Delta J}\,.
\eeq
Then, since operators at the exponent evaluated at different points commute, we can apply the usual cumulant expansion and find
\begin{equation}
\tilde{U}_{\Delta t} = \exp \left[{ \sum_{n=1}^\infty \frac{(-\Delta t)^n}{n!} \kappa_n(\omega)\, \partial^n_{\vartheta}} \right]\,,
\end{equation} 
where $\kappa_n$ is the $n$-th cumulant of the probability distribution of the frequencies $\omega$. The time evolution becomes
\begin{equation}
\tilde{f}_{t + \Delta t} = \exp\left[ \Delta t \, \partial_t \right] \tilde{f}_{t} = \exp \left[{ \sum_{n=1}^\infty \frac{(-\Delta t)^n}{n!} \kappa_n(\omega)\, \partial^n_{\vartheta}} \right] \tilde{f}_{t}\,,
\end{equation} 
so that
\beq
\pd{\tilde{f}}{t} = \sum^{\infty}_{n=1} (-1)^n \frac{\kappa_n (\omega)(\Delta t)^{n-1}}{n!} \mpd{\tilde{f}}{\vartheta}{n}\,.
\label{eq:diff}
\eeq
Let $\delta\omega = \omega - \overline{\omega}$, where 
$\overline{\omega} = \omega(J) = \kappa_1(\omega)$ is the average of the distribution of the frequencies $\omega$. Then $\kappa_1(\delta\omega) = 0$ and $\kappa_n (\delta \omega) = \kappa_n (\omega)$ for any $n > 1$, so that we can extract the first term from the sum in Eq.\ \eqref{eq:diff} and obtain 
\beq
\pd{\tilde{f}}{t} = - \omega(J) \pd{\tilde{f}}{\vartheta} + \sum^{\infty}_{n=2}  (-1)^n D_n (J)  \mpd{}{\vartheta}{n} \tilde{f}\,,
\label{vlasov_coarse_aa} 
\eeq
where we have introduced the diffusion coefficients
\beq
D_n (J) = \frac{(\Delta t)^{n-1}}{ n!}  \kappa_n (\delta \omega)\,.
\label{Dn}
\eeq
Equations \eqref{vlasov_coarse_aa} and \eqref{Dn} are written as such only at the time $t$ chosen to define the action-angle coordinates. However, we can rewrite the equation in a coordinate-independent way, by noting that $-\omega(J) \partial_{\vartheta} = \pb{H,\cdot}$, so that Eq.\ \eqref{vlasov_coarse_aa} is nothing but Eq.\ \eqref{Mostgeneral2} with the coefficients $\mu_n(H)$ explicitly given\footnote{Being, at a given time $t$, $H = H(J)$, the action variable $J$ is implicitly a function of $H$.} 
as 
\beq
\mu_n = D_n(J(H)) [\omega(J(H))]^{-n}\,. 
\eeq
Note that Eq.\ \eqref{Dn} implies that all the diffusion coefficients vanish if $\omega$ does not depend on $J$: in this case $\tilde{f}$ obeys the Vlasov equation as the fine-grained $f$ does. This is coherent with our picture, because no randomness is present if $\omega$ does not depend on $J$ and all the particles coherently drift in $\vartheta$ with the same frequency. Indeed, in the harmonic case where $H$ is linear in $J$ and $\omega$ is constant the motion can be described in terms of normal coordinates without any damping.  As already discussed, the even terms are those responsible for the breaking of the time-reversal symmetry; we can estimate their order of magnitude as 
\beq
D_{2n} \propto \frac{(\Delta J)^{2n} (\Delta t)^{2n-1}}{ 2^{2n} (2n+1)!}\,, 
\eeq 
while the odd coefficients are even more suppressed with $n$, being the distribution of $\delta\omega$ even at the leading order. Hence, as long as $\Delta J$ and $\Delta t$ are not too large, only the very first terms of the sum in Eq.\ \eqref{vlasov_coarse_aa} will give a non-negligible contribution. To the leading order in $\Delta J$ and $\Delta t$ 
the evolution equation for $\tilde{f}$ becomes a Fokker-Planck equation.  
Retaining only the lowest order terms in Eq.\ \eqref{vlasov_coarse_aa} we can write $\delta\omega(y) = \omega'(J)(y - J)$, where $y$ is a random action uniformly distributed between $J - \frac{1}{2}\Delta J$ and $J + \frac{1}{2}\Delta J$; then, $\delta\omega$ is uniformly distributed in the interval $[-\frac{1}{2}|\omega'(J)|\Delta J, \frac{1}{2}|\omega'(J)|\Delta J]$ so that, denoting by $D^0_n(J)$ the leading order approximation of the diffusion coefficients, we have 
\beq
D^0_2(J) = \frac{\Delta t}{2}\kappa_2(\delta \omega) = \frac{1}{24}\left[\omega'(J) \Delta J \right]^2 \Delta t
\eeq
and the lowest order truncation of Eq.\ \eqref{vlasov_coarse_aa} can be written as
\beq
\pd{\tilde{f}}{t} = -\omega(J) \pd{\tilde{f}}{\vartheta} + D^0_2 (J) \frac{\partial^2 \tilde{f}}{\partial \vartheta^2}\,,
\label{fokkerplanck}
\eeq    
where   
\beq
D^0_2 (J) = \Delta t\frac{[\omega'(J) \Delta J]^2}{24}\,, 
\label{D02}
\eeq
and $\omega' = d\omega/dJ$. Being $\mu_2 = D^0_2(J) \omega^{-2}$ and $(d\omega/dJ)^2 \omega^{-2} = (d\omega/dH)^2$, Eq.\ \eqref{fokkerplanck} can be cast in the covariant form
\beq
\pd{\tilde{f}}{t} =  \pb{H, \tilde{f}} + \frac{1}{24}\Delta t (\Delta J)^2 \pb{\omega(H),\pb{\omega(H),\tilde{f}}}\,,
\label{fokkerplanck_cov}
\eeq
that is a special case of Eq.\ \eqref{eq:general_lo}. Equation \eqref{fokkerplanck_cov} can be interpreted in the corresponding Langevin formalism (see Appendix \ref{app_Langevin}). Note that we have taken advantage of the existence of action-angle coordinates (at least at a given time) to derive our results, but then we have expressed them in a covariant fashion, so that they do not depend on the choice of coordinates. 

\section{Scaling of damping times} 
\label{sec_numerics}
According to Eq.\ \eqref{fokkerplanck}, the characteristic damping time of $\tilde{f}$ is
\beq
\tau = \left(D_2^0\right)^{-1}\,.
\label{tau}
\eeq
Let us now ask how $\tau$ depends on the coarse graining scale. In order for our coarse graining to be independent of the choice of the coordinates in phase space we have to define its scale in terms of phase space volumes (i.e., surfaces since $d=1$), that are invariant under canonical transformations. Let $\Delta\Gamma = \Delta J \Delta \vartheta$, so that we may assume that
$\Delta J \propto \sqrt{\Delta\Gamma}$ and $\Delta \vartheta \propto \sqrt{\Delta\Gamma}$. Equation \eqref{D02} contains $\Delta t$ instead of $\Delta \vartheta$, because of the way we have performed the symplectic coarse graining; however, $\Delta t \propto \Delta \vartheta \propto \sqrt{\Delta\Gamma}$, so that  Eqs.\ \eqref{D02} and \eqref{tau} imply $\tau \propto (\Delta\Gamma)^{-3/2}$. If $S$ is the smallest surface we want to probe, neglecting the dynamics occurring on scales smaller that $S$, we have to choose $\Delta\Gamma \approx S$, so that the damping time at scale $S$ obeys the scaling relation
\beq
\tau \propto S^{-3/2}\,.
\label{scaling}
\eeq   
The time given by \eqref{scaling} is the time after which the dynamics of the fine-grained $f$ has moved to scales smaller than $S$ in phase space, so that our coarse-grained description is no longer able to detect it.   
One way to probe different scales is to look at the Fourier components $f_{\mathbf{k}}$ of the distribution function in phase space, where $\mathbf{k} = (k_p,k_q)$, with $k_q$ and $k_p$ its components along any couple of canonical coordinates $(p,q)$: 
\beq
f({\mathbf{k}},t) = \int e^{i(k_p p + k_q q)} f(p,q,t)\, dp\, dq \,;
\label{eq:fk}
\eeq   
a given  $f_{\mathbf{k}}$ probes a strip in phase space of width proportional to $k^{-1}$, where $k = \left(k_p^2 + k_q^2\right)^{1/2}$. Therefore, to describe the evolution of $f_{\mathbf{k}}$ we have to choose $S \propto k^{-1}$, and we expect $f_{\mathbf{k}}$ to damp out on a time scale
\beq
\tau_{\mathbf{k}} \propto k^{3/2}\,.
\label{eq:scalinglaw}
\eeq 

\subsection{Numerical results}

To check the scaling law \eqref{eq:scalinglaw}, we solved the Vlasov equation for various one-dimensional models on an $N_{q}\times N_{p}$ grid with a time step $\Delta t$ until a maximum time $t = t_{\text{max}}$ 
using a semi-Lagrangian method \cite{DeBuyl:cpc2014,SonnendruckerEtAl:jcp1999}, computing the evolution of the $f_{\mathbf{k}}$ starting from nonstationary configurations, and 
always finding good agreement between Eq.\ \eqref{eq:scalinglaw} and numerics. Results are reported in the following subsections, while the description of the protocol to measure the damping times $\tau_{\mathbf{k}}$ is described in Appendix \ref{app_numerics}.

\subsubsection{HMF model}
\begin{figure}
\includegraphics[scale=.55]{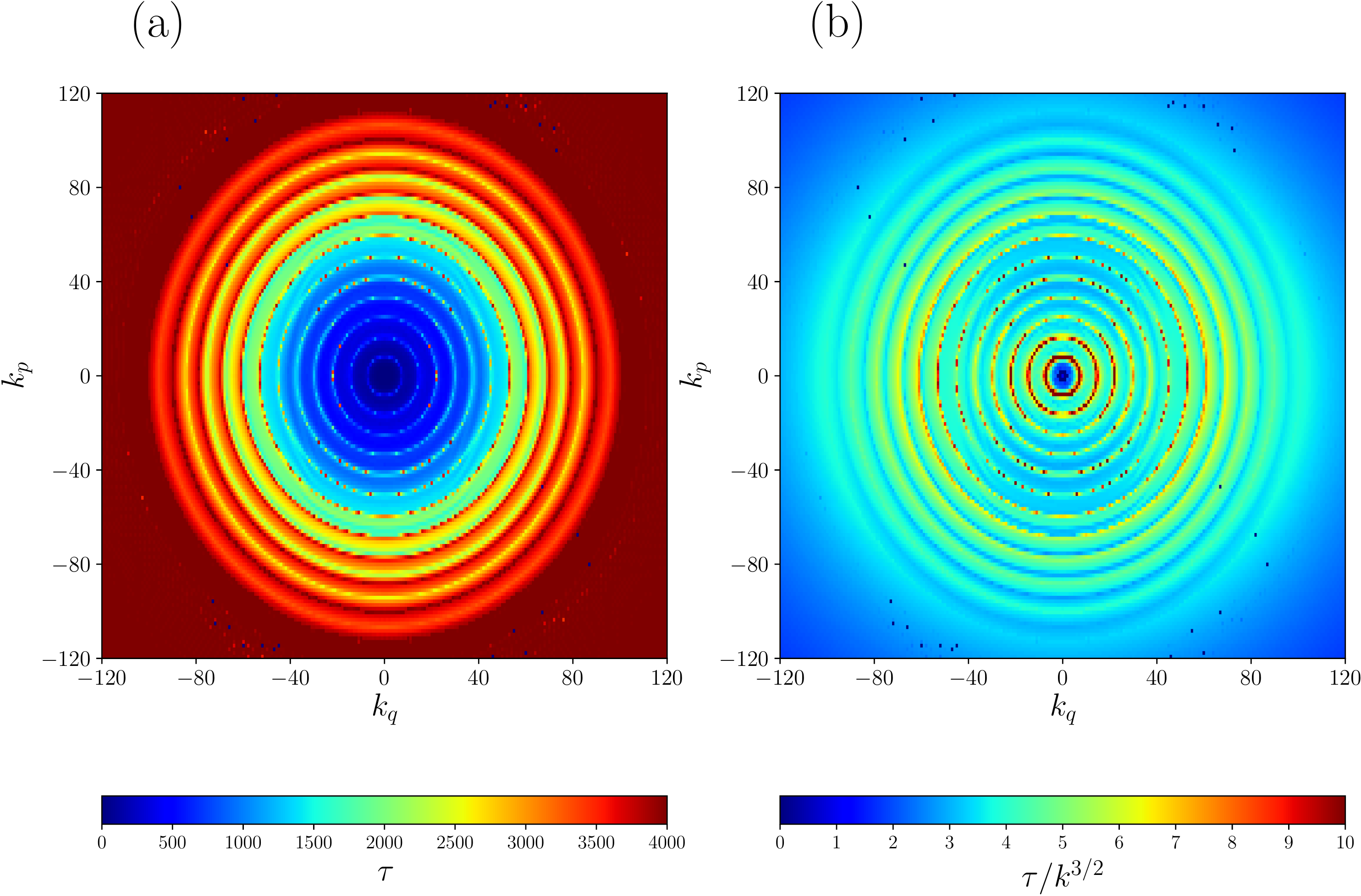}
\caption{HMF model: damping times of the Fourier components $f_{\mathbf{k}}$ of the distribution function defined in Eq.\ \eqref{eq:fk}. (a) Damping times as a function of ${\mathbf{k}}$. (b) Rescaled damping times $\tau_{\mathbf{k}}/k^{3/2}$. Note the difference in scale between the left and the right panel. The Vlasov equation was solved on an $N_{q}\times N_{p} = 1064 \times 1248$ grid with a time step $\Delta t = 2.5 \times 10^{-3}$ until $t = 4000$. Initial conditions: $q$'s and $p$'s  uniformly distributed in $[-\pi/2,\pi/2]$ and $[-0.25,0.25]$, respectively. 
}
\label{figura}
\end{figure}

\begin{figure}
\centering
\includegraphics[scale=.55]{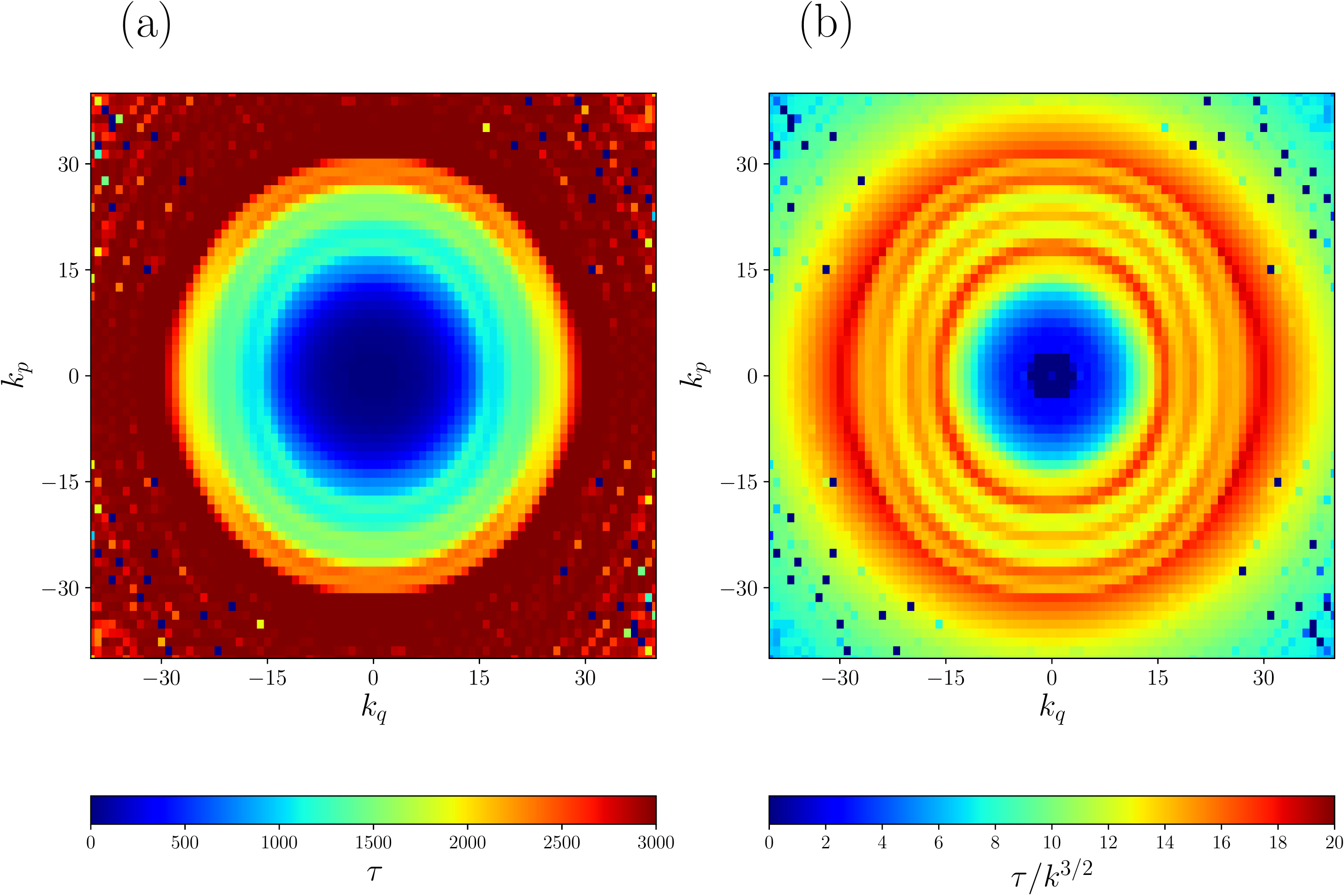}
\caption{As in Fig.\ \ref{figura}, with initial conditions such that the $q$'s are uniformly distributed in $[-1,1]$ and the $p$'s are normally distributed with zero mean and standard deviation equal to $0.1$; other simluation parameters are $N_{q}\times N_{p} = 1000 \times 1125$, $\Delta t = 2\times 10^{-3}$, and $t_{\text{max}} = 3\times 10^3$.}
\label{fig:HMF_gauss}
\end{figure}

The Hamiltonian Mean Field (HMF) model \cite{AntoniRuffo:pre1995} has been the workhorse of the studies on long-range-interacting systems in the last decades. The Hamiltonian of the model is
\begin{equation}
H = \sum_{i=1}^N \frac{p_i^2}{2} - \frac{J}{2N} \sum_{i=1}^N\sum_{j=1}^N \cos(q_i-q_j)\,,
\end{equation}
where $q_i \in [-\pi,\pi]$ and $p_i \in \mathbb{R}$, for $i = 1,\ldots,N$, are canonically conjugated coordinates. This model can be seen either as a system of globally coupled $XY$ spins or as $N$ particles with unit mass moving on a ring interacting via a cosine potential. In the following we shall use natural units to obtain dimensionless quantities, setting $J = 1$, thus considering only attractive (ferromagnetic, in the spin language) interactions. In the limit $N \to \infty$, the dynamics of the one-particle distribution function $f(q,p,t)$ is given by the Vlasov equation
\beq
\pd{f}{t} +p \pd{f}{q} - \td{V[f]}{q} \pd{f}{p} = 0 
\label{V:HMF}
\eeq
where
\begin{subequations}
\begin{align}
V[f](q) & = -m_x[f] \cos q - m_y[f] \sin q \,,\\
m_x[f] & = \int dq \, dp \, f \cos q \,,\\ 
m_y[f] & = \int dq \, dp\,  f \sin q\,. 
\end{align}
\end{subequations}
In Fig.\ \ref{figura} we show $\tau_{\mathbf{k}}$ as a function of $\mathbf{k}$ for $0 < k \lesssim 120$ and the same numbers rescaled according to Eq.\ \eqref{eq:scalinglaw}, for ``waterbag'' initial conditions, i.e., $q$'s and $p$'s drawn from a uniform distribution with compact support. While the $\tau_{\mathbf{k}}$'s span three orders of magnitude, almost all the rescaled times are $\mathcal{O}(1)$.
Damping times depend on the initial conditions: as an example, in Fig.\ \ref{fig:HMF_gauss} we show damping times and rescaled damping times obtained strating from different initial conditions with respect to the case shown in Fig.\ \ref{figura}: here, positions are still drawn from a uniform distribution with compact support, but now the momenta are drawn from a Gaussian distribution. The scaling law \eqref{eq:scalinglaw} is in good agreement with the data, although here the interval over which the rescaled damping times are distributed is larger than in the previous case (nonetheless, it is still $6 \times 10^{-3}$ times the interval of the values of the computed damping times). 

In the following we consider three other models living in one dimension: a one-dimensional scalar field interacting via a mean-field quartic potential\footnote{Note that the interaction in this model is not periodic in the coordinates, at variance with all the other models.}, a one-dimensional self-gravitating system, and the so-called self-gravitating ring (SGR) model. Being all the models one-dimensional, the Vlasov equation is of the form \eqref{V:HMF} for all of them, but the self-consistent potential energy $V[f(q)]$ will be different for each model. The initial conditions will be the same in all the examples we shall consider, and will be equal to those considered for the HMF model in the example reported in Fig.\ \ref{fig:HMF_gauss}, i.e., uniform on the segment $[-1,1]$ for the coordinates and Gaussian, with zero mean and standard deviation equal to 0.1, for the momenta. 

\subsubsection{Scalar field with mean-field quartic interaction}
\begin{figure}
\centering
\includegraphics[scale=.55]{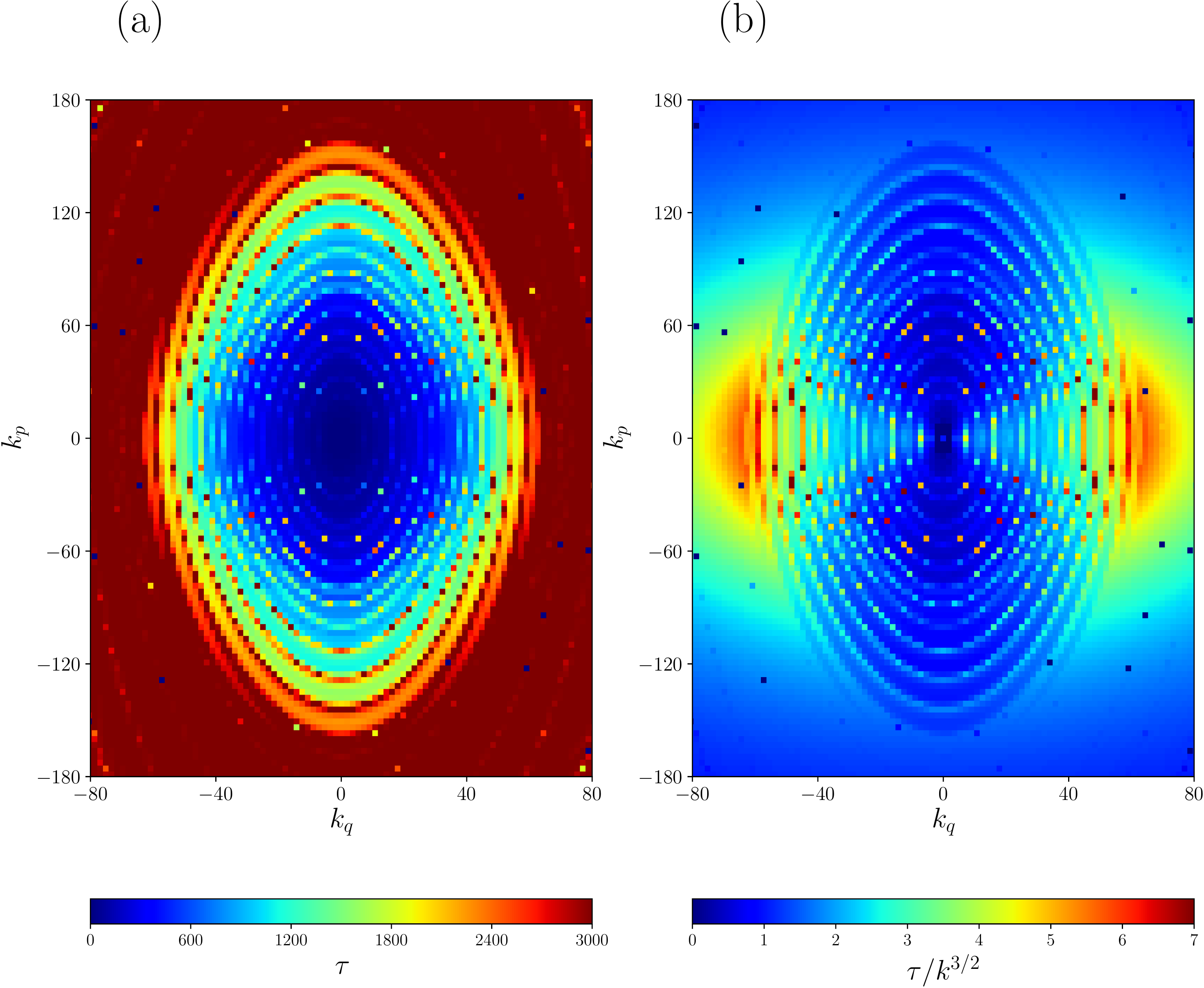}
\caption{As in Fig.\ \ref{figura}, for the scalar field with mean-field quartic interaction. Initial conditions are as in Fig.\ \ref{fig:HMF_gauss}, and other simulation parameters are $N_{q}\times N_{p} = 512 \times 1024$, $\Delta t =10^{-2}$, and $t_{\text{max}}= 3 \times 10^3$.}
\label{fig:Quartico_gauss}
\end{figure}

The mean-field-interacting scalar field model can be seen as the continuum limit of the $N$-particle Hamiltonian
\begin{equation}
H = \sum_{i=1}^N \frac{p_i^2}{2} + \frac{1}{2N} \sum_{i=1}^N \sum_{j=1}^N\frac{1}{4!} (q_i -q_j)^4~,
\end{equation}
where $q_i \in \mathbb{R}$ and $p_i \in \mathbb{R}$, for $i = 1,\ldots,N$, are canonically conjugated coordinates. 
In the Vlasov limit $N \to \infty$ the self-consistent potential energy  is
\beq
V[f](q) =  \int dq'dp' \frac{(q-q')^4}{4!} f(q',p',t) ~.
\eeq
An example of computed and rescaled damping times for this model is shown in Fig.\ \ref{fig:Quartico_gauss}. The agreement between the numerical data and the scaling law $\tau_{\mathbf{k}} \propto k^{3/2}$ is apparently very good. 

\subsubsection{One-dimensional self-gravitating system}
\label{sec:1dSG}
\begin{figure}
\centering
\includegraphics[scale=.55]{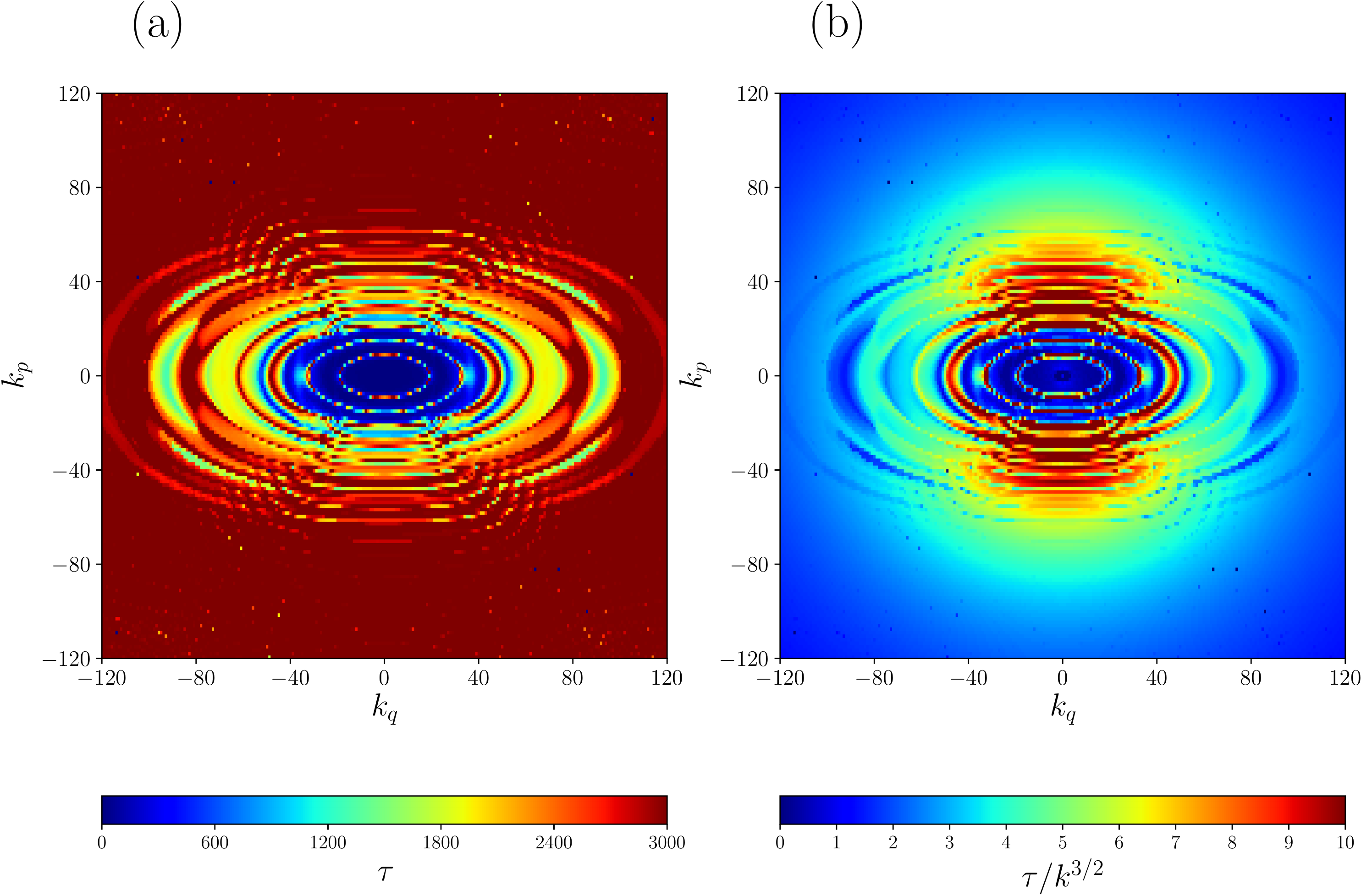}
\caption{As in Fig.\ \ref{figura}, for the one-dimensional self-gravitating system with periodic boundary conditions. Initial conditions are as in Fig.\ \ref{fig:HMF_gauss}, and other simulation parameters are $N_{q}\times N_{p} = 1024 \times 1024$, $\Delta t = 2\times 10^{-3}$ and $t_{\text{max}} = 3\times 10^3$.}
\label{fig:gravita_gauss}
\end{figure}

A one-dimensional self-gravitating system can be seen as $N$ infinite massive parallel planes, with a constant surface mass density, moving in the direction orthogonal to the planes themselves. Assuming periodic boundary conditions, the gravitational interaction can be expanded in a Fourier series, so that the Hamiltonian can be written, after introducing dimensionless variables, as 
\beq
H = \sum_{i=1}^N \frac{p_i^2}{2} -\frac{1}{2}\sum_{n=1}^\infty \frac{1}{n^2}\left( m_{x,n}^2 + m_{y,n}^2 \right)\,, 
\label{H:SG}
\eeq
where $p_i \in \mathbb{R}$, for $i = 1,\ldots,N$ and
\begin{subequations}
\begin{align}
m_{x,n} & = \frac{1}{N} \sum_{i=1}^N \cos\left(nq_i\right)\,, \\
m_{y,n} & = \frac{1}{N} \sum_{i=1}^N \sin\left(nq_i\right)\,, 
\end{align}
\end{subequations}
with $q_i \in [-\pi,\pi]$, for $i = 1,\ldots,N$. Hence the self-consistent potential entering the Vlasov equation for this model is
\beq
V[f](q) = -\sum_{n=1}^{+\infty} \frac{1}{n^2}\left[m_{x}^{(n)} \cos (nq) + m_{y}^{(n)} \sin (nq)\right] \,,
\label{V:SG}
\eeq
with
\begin{subequations}
\begin{align}
m_x^{(n)}[f] & = \int dq \, dp\,  f(p,q,t) \cos (nq)\,, \\
m_y^{(n)}[f] & = \int dq \, dp \, f(p,q,t) \sin (nq) \,.
\end{align}
\end{subequations}
In practice, one can consider a large but finite number $M$ of Fourier modes of the interaction, so that the infinite series in Eqs.\ \eqref{H:SG} and \eqref{V:SG} are replaced by finite sums with $n$ running from 1 to $M$; we considered $M = 250$. Note that if we take $M = 1$ we get back to the HMF model, whose interaction can then be seen as the lowest-order Fourier approximation of self-gravity in one dimension.  An example of computed and rescaled damping times for this model is shown in Fig.\ \ref{fig:gravita_gauss}. Again, the agreement between the numerical data and the scaling law \eqref{eq:scalinglaw} is very good. 

\subsubsection{Self-gravitating ring}
\begin{figure}[h]
\centering
\includegraphics[scale=.7]{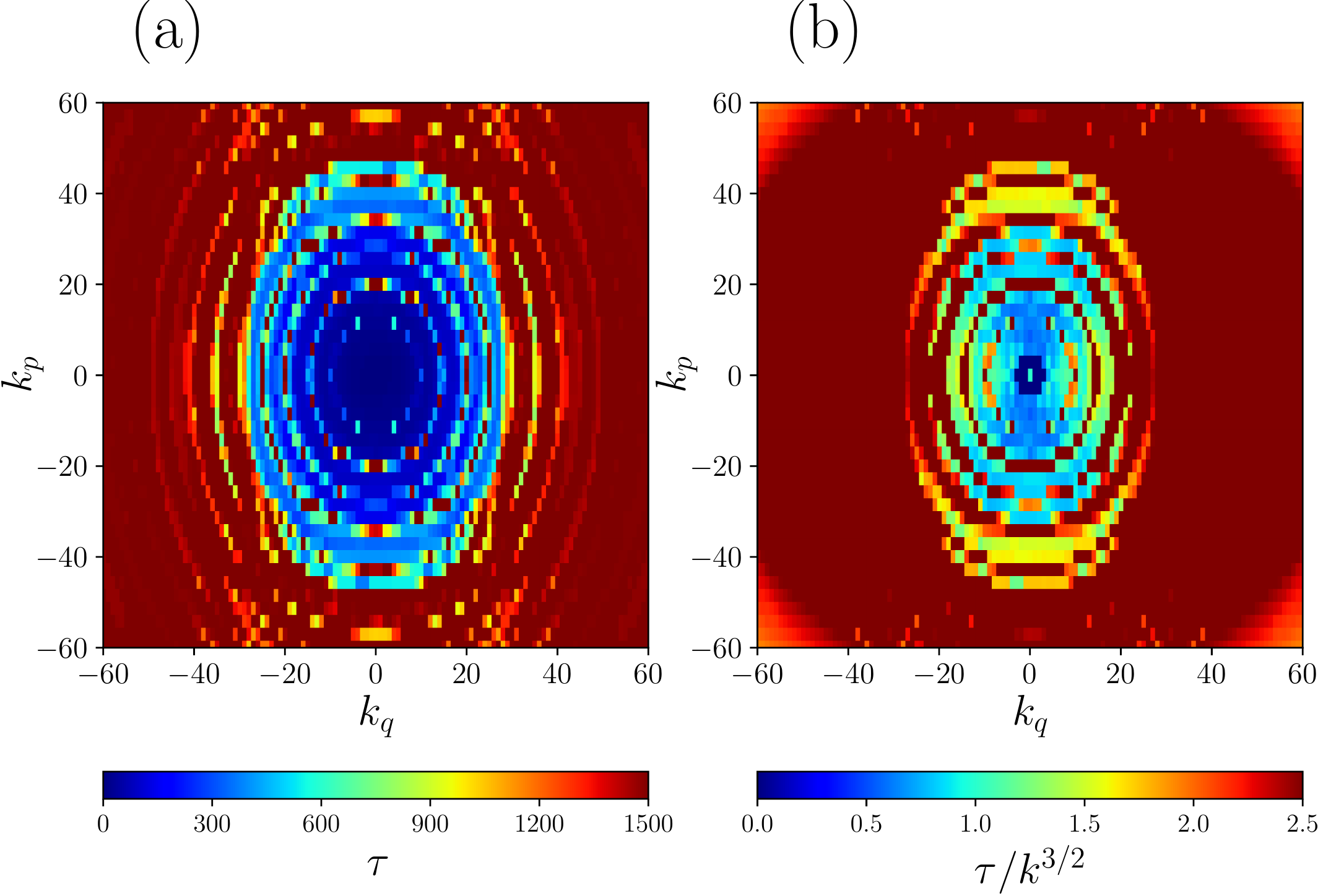}
\caption{As in Fig.\ \ref{figura}, for the SGR model. Initial conditions are as in Fig.\ \ref{fig:HMF_gauss}, and other simulation parameters are $N_{q}\times N_{p} =  700 \times 1024$, $\Delta t = 2.5\times 10^{-3}$ and $t_{\text{max}} = 1.5\times 10^3$.}
\label{fig:sgr_gauss}
\end{figure}

Instead of working as in Sec.\ \ref{sec:1dSG} with low-dimensional gravity,  one can consider (softened) three-dimensional gravitational forces but constrain the interacting particles to move on a ring; the resulting model is referred to as the Self-Gravitating Ring (SGR), introduced in  \cite{SotaEtAl:pre2001} and further studied in \cite{TatekawaEtAl:pre2005,prerap2009,jstat2010,pre2012_2}. The Hamiltonian, again expressed in dimensionless variables, is
\begin{equation}
H = \sum_{i=1}^N\frac{p_i^2}{2} -\frac{1}{2\sqrt{2}N} \sum_{i=1}^N \sum_{j=1}^N \left[\frac{1}{\sqrt{1-\cos(q_i-q_j) +\alpha}} \right],
\end{equation}
where $q_i \in [-\pi,\pi]$ and $p_i \in \mathbb{R}$, for $i = 1,\ldots,N$, are canonically conjugated coordinates and $\alpha$ is a softening parameter, regularizing the divergence of the gravitational interaction for vanishing distance between the particles. It can be shown \cite{TatekawaEtAl:pre2005} that the SGR reduces to the HMF in the limit $\alpha\to\infty$. The self-consistent potential entering the Vlasov equation for this model is written as
\beq
V[f](q) = - \frac{1}{\sqrt{2}} \int_{-\pi}^{\pi} dq'\int_{-\infty}^{\infty}dp' \frac{f(q',p',t) }{\sqrt{1-\cos(q-q')+\alpha}}~. 
\eeq
This model is somewhat harder to solve numerically than the previous ones, and numerical diffusion prevents reliable results for Fourier component $f_{\mathbf{k}}$ with large $k$'s, so that we had to limit ourselves to shorter simulations and to damping times corresponding to smaller wave vectors than in the previous cases. This notwithstanding, we are able to see the good agreement between the predicted scaling law $\tau_{\mathbf{k}} \propto k^{3/2}$ and the numerical results also for the SGR (see Fig.\ \ref{fig:sgr_gauss}).

\section{Conclusions} 
\label{sec_conclusions}
We have derived an effective evolution equation for a coarse-grained distribution function in the case of systems whose dynamics obeys the Vlasov equation in the $N\to\infty$ limit. A general form of the equation has been given based on symmetry considerations, i.e., requiring the conservation of the symplectic structure, and an explicit equation for 1-$d$ systems was derived independently of the general equation: the fact that we indeed found an equation of the same form of the general one is a nontrivial result and supports the validity of our approach. The lowest-order term of the equation is a diffusion along the Hamiltonian flow and becomes, if $f$ is stationary, a diffusion along the $J =$ constant lines in phase space. Diffusion in the stationary case (that implies a dynamics analogous to that dictated by a fixed external potential) is due to the dependence on $J$ of the frequency $\omega$, that entails differential rotation in phase space, a filamentation of $f$ and thus an effective mixing in phase space due to our blindness to small scales after coarse graining. Indeed, and as it should, if $\omega$ does non depend on $J$ as in the harmonic case no diffusion is present in our theory. Depending on $\omega$, diffusion may either be very effective (as it happens close to separatices) or not efficient at all. In the latter case our theory may predict long-standing oscillations, that may be an alternative outcome of violent relaxation instead of damping to a QSS \cite{TennysonEtAl:PhysicaD1994,BonifacioEtAl:RivNuovoCim1990,Mathur:MNRAS1990,Weinberg:ApJ1991}. Diffusion along equal action lines had already been shown to be effective for the HMF model in the stationary case \cite{LeonciniVanDenBergFanelli:epl2009}: the latter results are an independent, indirect check of the soundness of our approach. Our results provide a solid quantitative picture of the mechanism underlying violent relaxation  and shed light on the r\^{o}le of the coarse-graining scale: numerical results for 1-$d$ systems (here shown for the HMF model) are in very good agreement with our predictions. 

We note that an effective equation with the same kind of structure as the one presented here had been found for suitable moments of the distribution function, at the leading order and based on heuristic considerations, in \cite{jstat2019}. An effective description that gets rid of the small-scale Vlasov dynamics had already appeared in \cite{RobertSommeria:prl1992,ChavanisSommeria:prl1997}, where a phenomenological maximum entropy production principle was invoked to get a diffusion in velocity space, that apparently does not conserve the symplectic structure, at variance with our approach. In \cite{ChavanisBouchet:aa2005} a deterministic coarse-graining procedure was introduced, yielding a time-reversal-invariant effective evolution, at variance with the one we have derived here.   
As argued in \cite{BeraldoesilvaEtAl:apj2019}, a faster-than-collisional relaxation might also be induced by a finite number of particles $N$; however such mechanism seems not to be relevant to violent relaxation, that occurs in finite systems as well as in the Vlasov $N\to\infty$ limit, and whose timescale does not depend on $N$. Although we have explicitly derived the effective equation only in the 1-$d$ case, we expect the extension of our procedure to systems with $d > 1$ that are integrable at a given time, like e.g.\ the self-gravitating case with imposed spherical symmetry, to be possible. Moreover, one may think of applying a truncated form of the general equation, e.g., Eq.\ \eqref{eq:general_lo}, supplemented by some ansatz for the unknown function $\mu_2$, to describe violent relaxation  in generic long-range-interacting systems.

\acknowledgments

This work is part of MIUR-PRIN2017 project \textit{Coarse-grained description for non-equilibrium systems and transport phenomena (CO-NEST)} n.\ 201798CZL whose partial financial support is acknowledged. 

\appendix

\section{Proofs of some analytical results}

We present here proofs of some results put forward in the paper. First of all let us recall some properties of Poisson brackets. We consider a Hamiltonian system with $d$ degrees of freedom and Hamiltonian $H(q_1,\ldots,q_d,p_1,\ldots,p_d)$. Being by definition 
\begin{equation}
\pb{f, g} = \sum_{j = 1}^d \left( \pd{f}{p_j} \pd{g}{q_j} - \pd{f}{q_j} \pd{g}{p_j} \right) =  \sum_{j = 1}^d \pd{}{q_j} \left( \pd{f}{p_j} g \right) +\sum_{j = 1}^d \pd{}{p_j} \left( -\pd{f}{q_j} g \right) 
\end{equation}
that is a divergence in phase space, we have
\begin{equation} \label{prop1}
\int \pb{f,g} d \Gamma = 0\, ,
\end{equation}
for any $f$ and $g$ decaying sufficiently fast for large values of coordinates and momenta. In Eq.\ \eqref{prop1} the integral is extended to the whole $2d$-dimensional phase space and we have used the shorthand notation $d\Gamma = dp\, dq = \prod_{i=1}^d dp_i\, dq_i$ that we shall continue to use from now on. From Eq.\ \eqref{prop1}, considering three functions $f$, $g$ and $h$ again decaying sufficiently fast at infinity, and applying the Leibnitz rule
\begin{equation}
\pb{f, g h} = g \pb{f, h} +  h \pb{f, g}\,,
\end{equation}
we get the integration by parts formula 
\begin{equation} \label{prop2}
\int h \pb{f,g} d\Gamma = - \int g \pb{f,h} d\Gamma\, .
\end{equation}

\subsection{Conservation laws in the coarse-grained evolution} 
\label{app_conservationlaws}
\subsubsection{Conservation of the norm of $f$}
Using the identity 
\beq
\pb{M(H),\cdot} = M'(H) \pb{H,\cdot}
\label{eq:MRule}
\eeq
Eq.\ \eqref{Mostgeneral2} becomes
\beq
\pd{\tilde{f}}{t} =  \pb{H, \tilde{f}} + \sum^{\infty}_{n=2}  
\pb{M_n(H), \cdot}^{n} \tilde{f}\,,
\label{eq:M_n}
\eeq
where the $M_n$'s are such as $M'_n(H) = \mu_n(H)$. Integrating over the whole phase space and using Eq.\ \eqref{prop1} we have
\beq
\int \frac{\partial \tilde{f}}{\partial t} \, d\Gamma = 0\,,
\label{int_dfdt}
\eeq
and being $\int \partial_t \tilde{f} \, d\Gamma = \frac{d}{dt} \int \tilde{f}\, d\Gamma$ Eq.\ \eqref{int_dfdt} implies the conservation of the norm. 

\subsubsection{Conservation of the energy}
Working in one dimension to ease the notation, the energy of the system in a state defined by the coarse-grained distribution can be written as
\beq
E[\tilde{f}] = \frac{1}{2} \int p^2 \tilde{f}(p,q) \, d\Gamma + \frac{1}{2} \int \int \tilde{f}(p,q) V(q - q^{\prime}) \tilde{f}(p^{\prime},q^{\prime}) \, d\Gamma d\Gamma^{\prime} 
\eeq
leading to
\beq
\frac{\delta E}{\delta\tilde{f}} = \frac{p^2}{2} + \int V(q-q^{\prime}) \tilde{f}(q^{\prime}, p^{\prime})\, d \Gamma^{\prime} = \frac{p^2}{2} + \tilde{U}(q) \equiv H[\tilde{f}]\,.
\label{H[f]}
\eeq
The time derivative of the energy is
\beq
\frac{d}{dt}E[\tilde{f}] = \int \frac{\delta E}{\delta\tilde{f}} \frac{\partial\tilde{f}}{\partial t}\, d\Gamma\,,
\eeq
so that Eq.\ \eqref{H[f]} implies
\beq
\frac{d}{dt}E[\tilde{f}] = \int H[\tilde{f}] \frac{\partial\tilde{f}}{\partial t}\, d\Gamma\,.
\label{dEdt}
\eeq
Using Eq.\ \eqref{eq:M_n} we have
\beq
H\frac{\partial\tilde{f}}{\partial t} =  \frac{1}{2}\pb{H^2, \tilde{f}} + \sum^{\infty}_{n=2}  
\pb{\mathcal{M}_n(H), \cdot}^{n} \tilde{f}\,,
\eeq
where the $\mathcal{M}_n(H)$ are such that $\mathcal{M}'_n(H) = H\mu_n(H)$; integrating the above equation over the whole phase space and using Eq.\ \eqref{dEdt} we get $dE/dt = 0$. 

\subsection{Time evolution of convex Casimirs}

The fine-grained Vlasov evolution has infinite conserved quantities (Casimirs) obtained by integrating a generic function $C(f)$ of the distribution function $f$ over the whole phase space. Replacing $f$ by a coarse-grained one the Casimirs ar no longer constant of motion. However, among all the Casimirs defined using any coarse-graining distribution function $\tilde{f}$, that is
\beq
\mathcal{C}[\tilde{f}] = \int C(\tilde{f}) \, d\Gamma\,,
\eeq
those corresponding to a convex $C$ (that will be referred to as ``convex Casimirs'' from now on) must be non-increasing functions of time  \cite{TremaineHenonLynden-Bell:mnras1986}. In the case of one-dimensional systems, to be considered below, we are able to prove that our version of the coarse-grained dynamics does agree with such a constraint (see \S \ref{sec:convex1d} below). We did not succeed in proving such a result for the most general form \eqref{Mostgeneral2} of the coarse-grained evolution equation, but we can show that convex Casimirs do not increase with time if we restrict ourselves to the lowest-order truncation of Eq.\  \eqref{Mostgeneral2}.

\subsubsection{General case}
\label{app_convex_general}

Let us consider the lowest-order truncation of Eq.\  \eqref{Mostgeneral2}, that is 
\begin{equation} 
\label{truncation}
\pd{\tilde{f}}{t} =  \pb{H, \tilde{f}} +  \mu_2 (H) 
\pb{H, \pb{H,\tilde{f}}}\,,
\end{equation}
provided $\mu_2(H) \ge 0$. Indeed,
\beq
\frac{d}{dt} \mathcal{C}[\tilde{f}] = \int C'(\tilde{f}) \frac{\partial \tilde{f}}{\partial t} \, d\Gamma = 
\int C'(\tilde{f}) \pb{H, \tilde{f}} \, d\Gamma + \int \mu_2 (H)\, C'(\tilde{f}) 
\pb{H, \pb{H,\tilde{f}}} \, d\Gamma\,,
\eeq
and using Eq.\ \eqref{eq:MRule}, with  $\mu_2(H) = M^{\prime}_2 (H)$, we get
\beq
\frac{d}{dt} \mathcal{C}[\tilde{f}] = \int  \pb{H, C(\tilde{f})} \, d\Gamma + \int C'(\tilde{f})
\pb{M_2(H), \pb{H,\tilde{f}}} \, d\Gamma\,.
\eeq
The first integral in the r.h.s.\ of the above equation vanishes due to Eq.\ \eqref{prop1}, while integrating by parts the second term using Eq.\ \eqref{prop2} we get
\beq
\frac{d}{dt} \mathcal{C}[\tilde{f}] = -  \int \pb{M_2(H),C'(\tilde{f})} \pb{H,\tilde{f}} \, d\Gamma = - \int \mu_2(H)\, C''(\tilde{f})  \pb{H,\tilde{f}}^2 d\Gamma\,;
\label{eq:dCdt}
\eeq
being $C(\tilde{f})$ convex, this implies $\dot{\mathcal{C}}[\tilde{f}] \le 0$ provided $\mu_2(H) \ge 0$.  It is interesting to note that  Eq.\ \eqref{eq:dCdt} tells us that $\mathcal{C}[\tilde{f}]$ does not reach its minimum: its evolution eventually stops when $\tilde{f}$ approaches a stationary solution, that is, such that $\pb{H,\tilde{f}} = 0$. The latter is a necessary feature of a consistent evolution, because it would not be possible, in general, to reach a state where all the (infinite) convex Casimirs are simultaneously minimized (see also the discussion in Ref.\ \cite{TremaineHenonLynden-Bell:mnras1986}).  

\subsubsection{One-dimensional systems}
\label{sec:convex1d}
Let us now show that the evolution defined by Eqs.\ \eqref{eq:evolution1d} and \eqref{eq:evol_operator} fulfills the constraint on the evolution of convex Casimirs. To this end we explicitly write down the average in Eq.\ \eqref{eq:evol_operator} in terms of action-angle variables at time $t$,
\beq
\left\langle e^{\Delta t \{H,\cdot \} }\right\rangle_{\Delta\Gamma} = \frac{1}{\Delta J \Delta\vartheta}\int_{\Delta J} dJ' \int_{\Delta \vartheta} d\vartheta' e^{- \omega(J') \Delta t \partial_{\vartheta}} = \frac{1}{\Delta J}\int_{\Delta J} dJ' \, e^{- \omega(J') \Delta t \partial_{\vartheta}}~,
\eeq
where we dropped the average over the angle variable since the integrand only depends on $J^{\prime}$. Considering now a sufficiently small $\Delta t$ such as to expand the exponential in the above equation up to first order and applying the resulting evolution operator to the coarse-grained distribution function evaluated at time $t$, $\tilde{f}_t$, we get 
\begin{equation}
\tilde{f}_{t+\Delta t} (\vartheta,J) = \frac{1}{\Delta J}\int_{\Delta J} {dJ^{\prime}}\, \left[ \tilde{f}_t - \omega(J') \Delta t \pd{\tilde{f}_t}{\vartheta}\right]\,.
\label{eq:ft+dt}
\eeq
To first order accuracy we can replace the derivative in the above equation with the ratio of $\tilde{f}(\vartheta,J) - \tilde{f}(\vartheta -\omega(J')\Delta t ,J)$ and $\omega(J')\Delta t$, so that Eq.\ \eqref{eq:ft+dt} becomes 
\beq
\tilde{f}_{t+\Delta t} (\vartheta,J) = \me{ \tilde{f} \left( \vartheta - \omega(J^{\prime}) \Delta t, J \right)}_{\Delta J}\,.
\end{equation}
On the other hand, for any convex function $C(x)$ and for any random variable $x$, 
\begin{equation}
C(\me{x}) \leq \me{C(x)}\,,
\end{equation}
so that 
\begin{equation}
C\left( \tilde{f}_{t+\Delta t} \right) \leq \me{ C \left( \tilde{f}\left( \vartheta - \omega(J) \Delta t, J \right) \right)}_{\Delta J}\,,
\end{equation}
or, explicitly writing the average over $\Delta J$ once again,
\begin{equation}
C \left( \tilde{f}_{t+\Delta t} \right) \leq \frac{1}{\Delta J}\int_{\Delta J} d J'\,  C \left( \tilde{f} \left( \vartheta - \omega(J') \Delta t, J \right) \right)\,.
\end{equation}
To obtain a condition on the Casimir functional at time $t + \Delta t$ we have to integrate the above relation in $\vartheta$ and $J$ all over the phase space, obtaining
\beq
\mathcal{C} \left[ \tilde{f}_{t+\Delta \tau} \right] \leq \frac{1}{\Delta J}\int_{\Delta J} d J' \int dJ \int d\vartheta \, C \left( \tilde{f} \left( \vartheta - \omega(J') \Delta t, J \right) \right)\,,
\label{diseq_int}
\eeq
but $\tilde{f}$ is a periodic function of $\vartheta$, so that
\beq
\int d\vartheta \,  C \left( \tilde{f} \left( \vartheta - \omega(J') \Delta t, J \right) \right) = \int d\vartheta  \, C \left( \tilde{f} \left( \vartheta,  J \right) \right)
\eeq
that no longer depends on $J'$; the average over $\Delta J$ is thus trivial and Eq.\ \eqref{diseq_int} becomes
\begin{equation}
\mathcal{C} \left[ \tilde{f}_{t+\Delta \tau} \right] \leq \mathcal{C} \left[ \tilde{f}_{t} \right],
\end{equation}
that is, what we wanted to prove.

\section{Langevin equation}
\label{app_Langevin}

It is interesting to note that the leading-order effective evolution equation in the one-dimensional case \eqref{fokkerplanck_cov}
can be cast in the form of a Fokker-Planck equation and interpreted, in turn, in the corresponding Langevin formalism. The nested Poisson bracket in Eq.\ \eqref{fokkerplanck_cov} can be written as
\beq
\pb{\omega(H),\pb{\omega(H),\tilde{f}}} = \partial_{x_i} \left( v_i \partial_{x_j} \left( v_j f \right) \right)\,, 
\eeq
where $i = 1,2$, $x_1 = q$, $x_2 = p$, $v_i = \varepsilon_{ij} \partial_{x_i} \omega$ with $\varepsilon_{ij}$ the totally antisymmetric Levi-Civita symbol and we have used the Einstein summation convention over repeated indices, so that Eq.\ \eqref{fokkerplanck_cov} becomes
\begin{equation}
\pd{\tilde{f}}{t} = - \varepsilon_{ij} \partial_{x_i} \left( \partial_{x_j} H \tilde{f}\right) +  \frac{1}{24} \Delta t (\Delta J)^2 \partial_{x_i} \left( v_i \partial_{x_j} \left( v_j \tilde{f} \right) \right) \,,
\end{equation}
in which we recognize the general form of a Fokker-Planck equation with a non-isotropic and non-uniform diffusion coefficient. Such an equation is in turn equivalent, in the Langevin formalism, to the Stratonovich differential equation \cite{vanKampen:book}
\begin{equation}
\dot{x_i} = \varepsilon_{ij} \partial_{x_j} H +  v_j \xi(t)
\end{equation}
$\xi(t)$ being a white noise with correlation function
\begin{equation}
\me{\xi (t) \xi(t')}= \frac{1}{24} \Delta J^2 \Delta t \, \delta(t-t') \, .
\end{equation}
Exploiting the definition of $v_j$ we can write
\begin{subequations}
\begin{align}
\dot{q} &= \pd{H}{p} + \pd{\omega (H)}{p} \xi(t)\,, \\
\dot{p} &= - \pd{H}{q} - \pd{\omega (H)}{q} \xi(t)\,;
\end{align}  
\end{subequations}
as expected, this couple of equations can be derived from the stochastic Hamiltonian
\begin{equation}
\tilde{H} = H + \omega (H) \, \xi(t)
\end{equation}
where once again we are using the Stratonovich formalism. 

\section{Measuring damping times in numerical simulations}
\label{app_numerics}

\begin{figure}
\centering
\includegraphics[scale=.6]{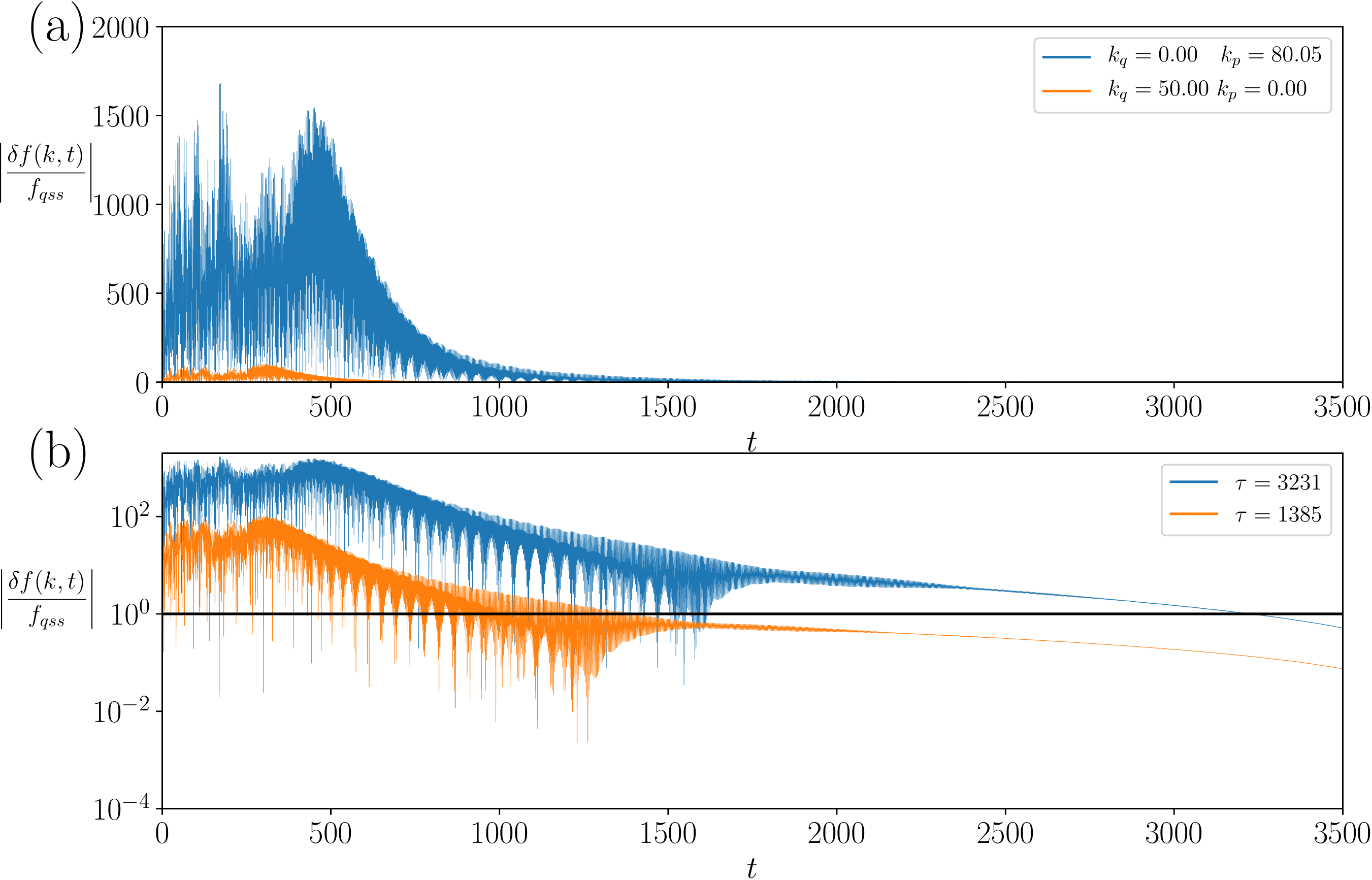}
\caption{HMF model. (a) Examples of $[f(\mathbf{k},t) - f^{\text{qss}}(\mathbf{k})]/ f^{\text{qss}}(\mathbf{k})$ (see legend for the values of $\mathbf{k}$), in linear scale. (b) As in (a) but in log-linear scale. The horizontal black line defines our threshold (equal to 1 here); the damping time $\tau$ is such that the curve is below the horizontal line for any $t > \tau$, and the values of $\tau$ obtained for the two cases plotted in the figure are reported in the legend. Simulation parameters are as in Fig.\ \ref{figura}, i.e., $N_{q}\times N_{p} = 1064 \times 1248$, $\Delta t = 2.5 \times 10^{-3}$, and $t_{\text{max}} = 4\times 10^3$.}
\label{fig:HMF_details}
\end{figure}

\begin{figure}
\centering
\includegraphics[scale=0.195]{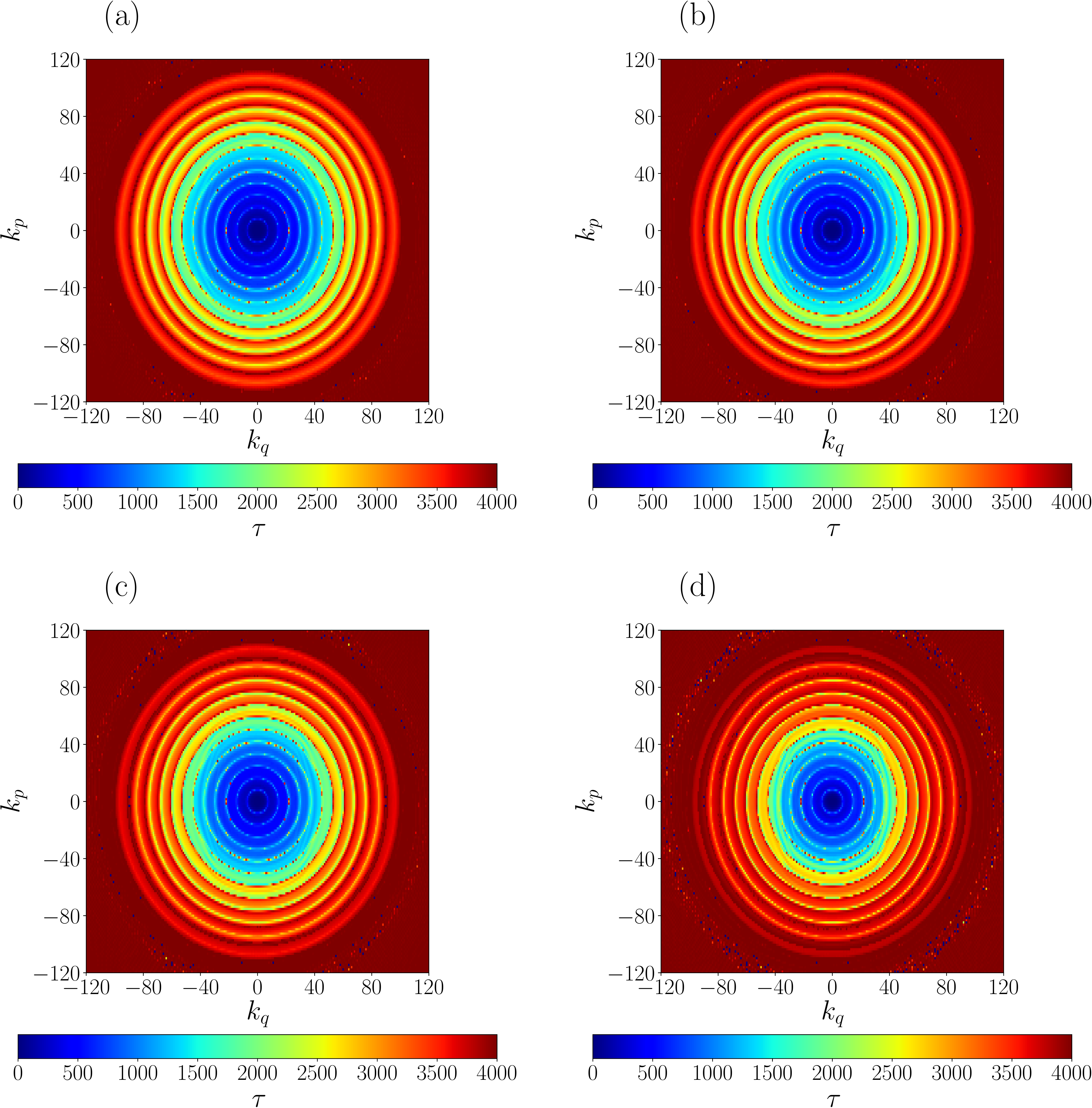}
\includegraphics[scale=0.193]{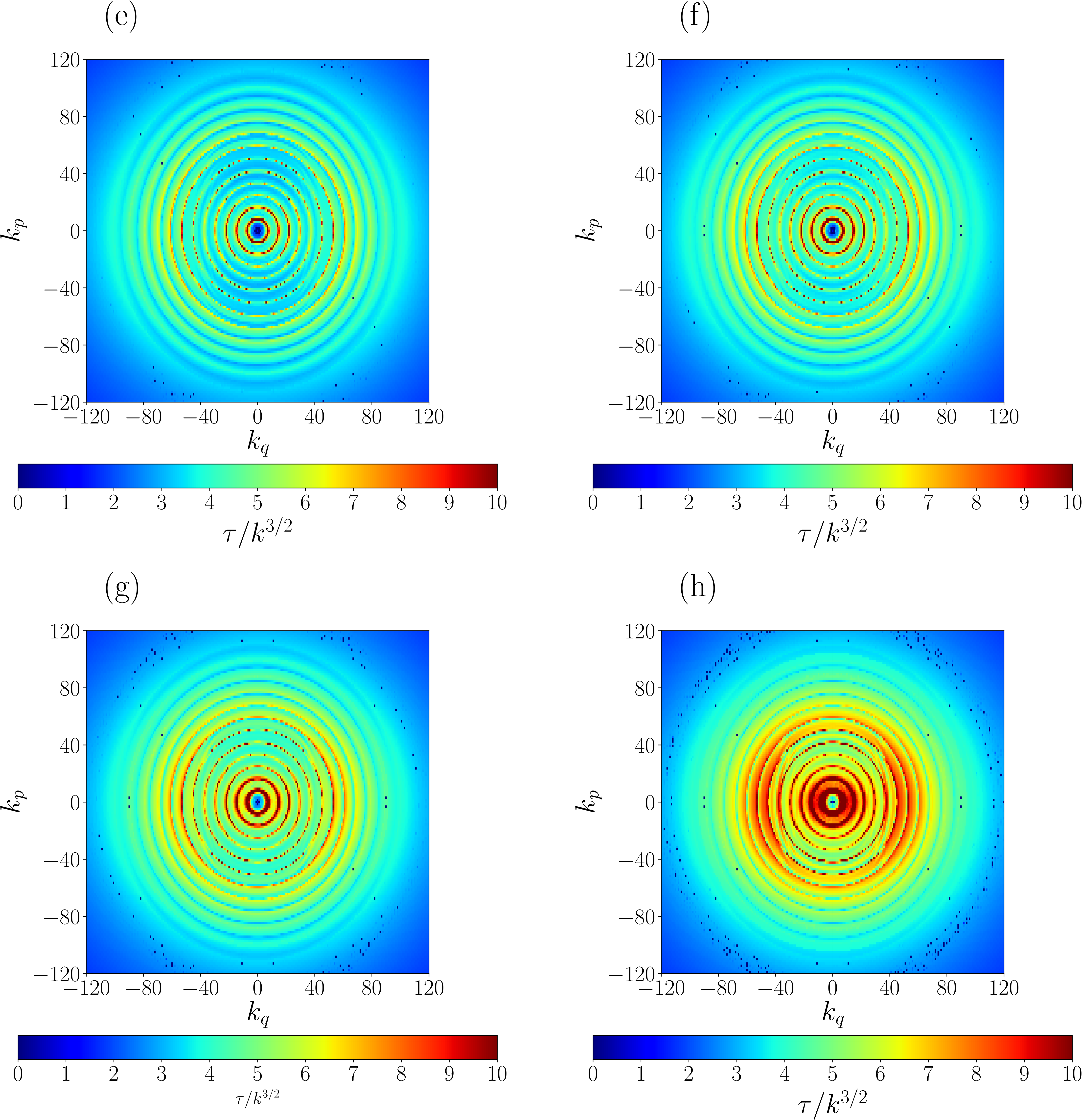}
\caption{Dependence of computed and rescaled damping times on the threshold used to define the damping time---see Eq.\ \eqref{cutoff}---for the HMF model. We show here how the results reported in Fig.\ \ref{figura} are affected by the choice of a different threshold: in (a), (b), (c) and (d) we plot the damping times $\tau$ as a function of $\mathbf{k}$ for four different choices of the threshold. Values of the threshold are (a) 1.0, (b) 0.75, (c) 0.50, (d) 0.25. In (e), (f), (g) and (h) we plot the corresponding rescaled damping times $\tau(\mathbf{k})/k^{3/2}$: values of the threshold are (e) 1.0, (f) 0.75, (g) 0.50, (h) 0.25. Note the change of scale between (a)-(d) end (e)-(h). Simulation parameters as in Fig.\ \ref{fig:HMF_details}.}
\label{fig:cutoff}
\end{figure}

 We defined the damping time $\tau_{\mathbf{k}}$ as the time for which the deviation of $f({\mathbf{k}},t)$ from its asymptotic value $f^{\text{qss}}(\mathbf{k})$ is definitively smaller than $f^{\text{qss}}(\mathbf{k})$ itself, that is, $\tau_{\mathbf{k}}$ is such as 
\begin{equation}
\left|\frac{\delta f(\mathbf{k},t)}{f^{\text{qss}}(\mathbf{k})}\right|\equiv\left|\frac{ f(\mathbf{k},t) - f^{\text{qss}}(\mathbf{k})}{f^{\text{qss}}(\mathbf{k})}\right|< 1\qquad \forall\, t \geq \tau_{\mathbf{k}}\,.
\label{cutoff}
\end{equation}
The asymptotic value $f^{\text{qss}}(\mathbf{k})$ is defined as the average of $f({\mathbf{k}},t)$ over the final part of the simulation, of duration $t_0 = 250$. In Fig.\ \ref{fig:HMF_details} we report the time evolution of $[f(\mathbf{k},t) - f^{\text{qss}}(\mathbf{k})]/ f^{\text{qss}}(\mathbf{k})$ for two particular Fourier components (extracted from the simulation used to obtain the results shown in Fig.\ \ref{figura}) to clarify the definition of the damping times.
The threshold we used, that is, the fact that the r.h.s.\ of the inequality in Eq.\ \eqref{cutoff} equals 1, is somewhat arbitrary, and any other number not so far from unity would make sense.  
For this reason we show in figure \ref{fig:cutoff} how the results presented in Fig.\ \ref{figura} are affected by the choice of different threshold values. It is apparent that a smaller threshold implies longer damping times, but the damping times still follow the scaling $\tau_{\mathbf{k}} \propto k^{3/2}$ with more or less the same accuracy (maybe getting only slightly worse for smaller threshold) for any choice of the threshold. In all the results presented in the paper the threshold has been kept equal to 1 as in Eq.\ \eqref{cutoff}.

\input{vlasov_renorm_transfer_newfigs.bbl}



\end{document}

%% file: vlasov_renorm_transfer_newfigs.bbl
%